\documentclass[useAMS, usenatbib]{mnras}
\usepackage{amsmath}
\usepackage{mathtools}
\usepackage{cases}
\usepackage{graphicx}
\usepackage{bigstrut}
\usepackage{enumitem}
\usepackage{graphicx}
\usepackage{paralist}
\usepackage{lscape}
\usepackage{wrapfig}
\usepackage[section]{placeins}
\usepackage[T1]{fontenc}
\usepackage{aecompl}

\newcounter{para}

\title[Monitoring for the optical variability in 3C\,66A]{Optical intra-day variability in  3C\,66A: 10 years of observations}

\author[Navpreet Kaur et al.]
 {Navpreet Kaur$^{1, 2}$,  Sameer$^{1,3}$, 
Kiran S. Baliyan$^{1}$\thanks{E-mail: baliyan@prl.res.in(KSB:Corresponding author)},
 and S. Ganesh$^{1}$ 
 \\ 
 $^{1}$Astronomy \& Astrophysics Division, Physical Research Laboratory, Ahmedabad 380009, India\\
 $^{2}$Indian Institute of Technology, Gandhinagar 382355, Gujarat,  India\\
$^3$ Department of Astronomy \& Astrophysics, Davey Laboratory, PSU, PA 16802 USA 
 }

\begin{document}
\bibliographystyle{mnras}
\date{Accepted for publication in MNRAS}

\maketitle

\label{firstpage}

\begin{abstract}

We present results based on the observations of the blazar 3C\,66A from 2005 November 06 to 2016 February 14  in the  BVR and I broadbands using 1.2m telescope of the Mt. Abu InfraRed Observatory (MIRO). The source was observed on 160 nights out of which on  89 nights it was monitored for more than 1 hr to check for the presence of any  intra-day variability (IDV). The blazar 3C\,66A exhibited significant variations in the optical flux on short and long term time scales. However, unlike highly variable S5 0716+71, it showed  IDV duty cycle of about 8\% only. Our statistical studies suggest the IDV time scales ranging from $\sim$ 37 min to about 3.12 hours and, in one case, a possibility of the quasi-periodic variations with characteristic timescale of $\sim$ 1.4 hrs. The IDV amplitudes in R$-$band were found to vary from 0.03 mag to as much as 0.6 mag, with larger amplitude of variation when the source was relatively fainter. The typical rate of the flux variation was estimated to be $\sim$0.07 mag hr$^{-1}$ in both, the rising and the falling phases. However, the rates of the brightness variation as high as 1.38 mag/hr were also detected. The shortest timescale of the variation, 37 min, sets an upper limit of $6.92 \times 10^{14}$ cm on the size of the emission region and  about $3.7 \times 10^8\ \mathrm{M}_{\odot}$ as an estimate of the mass of the black hole, assuming the origin of the rapid optical variability is in close vicinity of the central SMBH. The long-term study suggests a mild bluer-when-brighter behavior, typical for BL Lacs. 

\end{abstract}

\begin{keywords}
galaxies: active -- BL Lacertae objects : general --BL Lacertae objects : individual (3C\,66A) -- techniques : photometry--method: observational.
\end{keywords}

\section{Introduction}

3C\,66A was first identified by \citet{Wills1974} as a blue stellar object having a featureless optical spectrum with (U-B) and (B-V) colors similar to those found for the quasi-stellar objects. It is classified as an intermediate BL Lac (IBL) as its synchrotron emission component usually peaks between $10^{14}$ Hz and $10^{15}$ Hz \citep{Perri2003, Abdo2010b}. An early investigation of this object by \citet{Folsom1976} demonstrated that this object was variable in the optical with flare amplitude of approximately 0.5 mag and time scale of variation ranging from a few days to months. The WEBT campaign of 2003-2004 \citep{Bottcher2005} revealed micro-variability with flux changes of $\sim$5\%\ on a timescale of $\sim$ 2hr.  Intra-day variability (IDV) has also been addressed by several other authors \citep[e.g.][]{ Gopal2011, rani2011, Sagar2004, Raiteri1998} albeit with limited number of nights over a maximum  period of 4-years. \citet{Williamson2014} used X-ray, $\gamma $-ray and optical data during 2008-2012 to discuss spectral behavior in flaring and quiescent states of 3C\,66A \citep[also see][]{abdo20113c66a} and other blazars. \citet{Bach2007}, while studying a larger sample of blazars, found 3C\,66A to be variable at months time scale with faster variations superposed on its optical light-curve during 2005-2006.  However, long term behavior of this source has not been studied to explore its structure and any relationship between the rapid variations  and the long term trends using a consistent data-set. Such long term studies on some of the  blazars have resulted in detection of, for example,  the  presence of the quasi-periodic variations in OJ287, explained based on the binary black hole model  \citep{Sillanppa1988, pihajoki2013} and the possibility of the precession of the jet \citep{nesci2005} in S5 0716+714. \citet{Romero1999} suggested that the long term variability (LTV) arises due to the large scale relativistic shock moving down the jet, while IDV was caused by the  interaction of these shocks with small scale particle or magnetic field irregularities present in the jet. If  both, IDV and STV/LTV, features are related to the processes in the jet then a long term high temporal resolution study of the blazars would be very useful to see if there was any connection between the them.

\smallskip
In the present work, we use long-term  optical monitoring of the blazar 3C\,66A during the period 2005-2016 to discuss occurrence of the IDV and the behavior of the source at longer time scales. It would also be interesting to see if the short term variations play any role in the variability at longer time scales. Next section discusses observations and the data analysis, section 3 deals with the results and discussion and  the last section provides a summary of the work. 

\section{Observations and Data Analysis}
The photometric monitoring of 3C\,66A was performed using the PIXELLANT liquid Nitrogen (LN2) cooled CCD-Camera mounted at the f/13 Cassegrain focus of the 1.2m Telescope at Mt. Abu InfraRed Observatory (MIRO), Mt Abu, Rajasthan, India. The CCD Camera has 22 micron $1296 \times 1152$ pixels and a  read out time of about 13 seconds. With  0.29 arcsec per pixel, the total field of view (FOV) is about  $6.5' \times 5.5'$.  The observations during October 2014 were made with thermo-electrically cooled  iKon ANDOR CCD camera having $2048 \times 2048$ pixels. The CCD-photometric system is equipped with the Johnson-Cousin UBVRI \citep{johnson1951, bessell2005} filter set. The general observation strategy was to take about 4 frames in B, V, R and I bands and monitor the source in the R-band for a few hours. During our observation campaign from 2005 November 06 to 2016 February 14, the blazar was observed for a total of 160 nights.  A total of 15173, 702, 1187 and 597 photometric data points in R, B, V and  I bands, respectively, were obtained during more than ten years of observations.

\smallskip
Data reduction is performed using standard routines in IRAF\footnote{IRAF is distributed by the National Optical Astronomy Observatories, which are operated by the Association of Universities for Research in Astronomy, Inc., under cooperative agreement with the National Science Foundation.}. The instrumental magnitudes are obtained via aperture photometry carried out with several aperture sizes, ranging from 1 to 5 times the FWHM, and using the one with size 3 times the FWHM which led to optimum value of the S/N  as prescribed by \citet{Cellone2000}.

Calibration of the source magnitude is done by differential photometry with respect to the comparison stars in the same field, which reduces the effect of changes in the atmospheric seeing. In order to obtain a reliable photometric sequence for 3C\,66A we selected a set of non-variable stars with brightness closer to that of the object.  For each CCD frame, the instrumental magnitudes of the blazar 3C\,66A and five other stars were extracted. Out of the five stars, four were found appropriate to be used as standard comparison stars viz.  1, 2, 4 and 5\footnote{http://www.lsw.uni-heidelberg.de/projects/extragalactic/charts}.  The photometric sequence for these stars was reported by \cite{smith1998, fiorucci1996,  craine1975} (here-on referred to as C1, C2, C4 and C5, respectively). We used one of the four stars closest in brightness to the source, as a comparison star and rest of the available ones as control stars to check the stability of the sky. The observed magnitudes of the 3C\,66A obtained during the non-IDV nights were corrected using the averaged magnitude of all the four stars. Since the color difference between 3C\,66A and its comparison stars is low ($\sim$ 0.5 mag), the effect of the air mass variation has negligible impact on our photometric results \citep{carini1992}.

\section{Results \& Discussion}
The instrumental magnitudes obtained from the aperture photometry carried out  on the source and the field stars of interest  are  used to generate differential light-curves for the source-comparison and the control-comparison stars.  The standard values of all the comparison/control stars are used to calibrate the source magnitudes. The C-Parameter test \citep{Jang&Miller1997, Romero1999} and F-test (ratio of sample variances) techniques, widely used \citep[][and references there-in]{chandra2011, rani2011} to detect and confirm intra-day variations are used to calculate variability parameters, e.g., time scale and amplitude of variation etc. In this section we discuss intra-day variability using the differential light-curves as obtained for all the IDV qualified nights. Also, the daily averaged brightness magnitudes for the source 3C\,66A for all the nights and in all the filters are used to generate long term light-curves and color-magnitude diagrams to study long term behavior of the source.

\begin{table*}
\caption{\bf Details of the confirmed IDV nights for 3C\,66A with  no. of data points (N), duration of observation,  averaged  R-band magnitudes, variability parameter (C), amplitude of variation ($A_{var}$), F-test value, slope of structure function (SF Slope) and variability time scale ($t_{var}$).\label{tab:table1}}
\textwidth=7.0in
\textheight=10.0in
\vspace*{0.2in}
\noindent  
\begin{tabular}{cccccccccccc} \hline \nonumber
Date             & MJD              & N         & Duration  & R-band & $\sigma_{R}$  & C 		 & $A_{var}$  & F-test     	 & SF Slope   				& $t_{var}$  
\\
(yyyy-mm-dd)	 	   & (start)	   	   &	          & (mins) & 	(mag)	&  &		 &(mag)		      &               	 & ($\beta$) 				& (mins)
\\
\hline 

2005-12-23  & 53727.9075  & 127	& 194.85	    & 14.49		&0.01      & 3.26	 & 0.14	       & 10.63	  & 0.76$\pm$0.03	 	&  118    
\\ 
2006-11-16  & 54055.0058  & 93	& 85.90        & 14.49		&0.01      & 4.43	 & 0.11	       & 19.62	  & 1.96$\pm$0.07		&  68
\\  
2009-11-21  & 55156.9100  & 103	& 185.95      & 14.71		&0.01      & 4.87	 & 0.67	       & 23.72	  & 1.19$\pm$0.03		&  37
\\ 
2010-11-07  & 55507.0109  & 139	& 190.35      & 13.86		&0.01      & 4.62	 & 0.32	       & 21.34	  & 1.40$\pm$0.03		& 187
\\ 
2012-11-21  & 56252.8389  & 144	& 226.73      & 14.40		&0.01     & 3.56	  	& 0.17	       & 12.67	  & 1.35$\pm$0.05		& 40
\\ 
2015-12-30  & 57386.6011  & 198	& 213.74	   & 14.65		&0.01     & 3.16	 	& 0.15	       & 9.98	  & 1.41$\pm$0.05		& 41
\\
 \hline 
     \end{tabular} \\
\end{table*}

\begin{figure*} 
 \includegraphics[width = 0.3\textwidth]{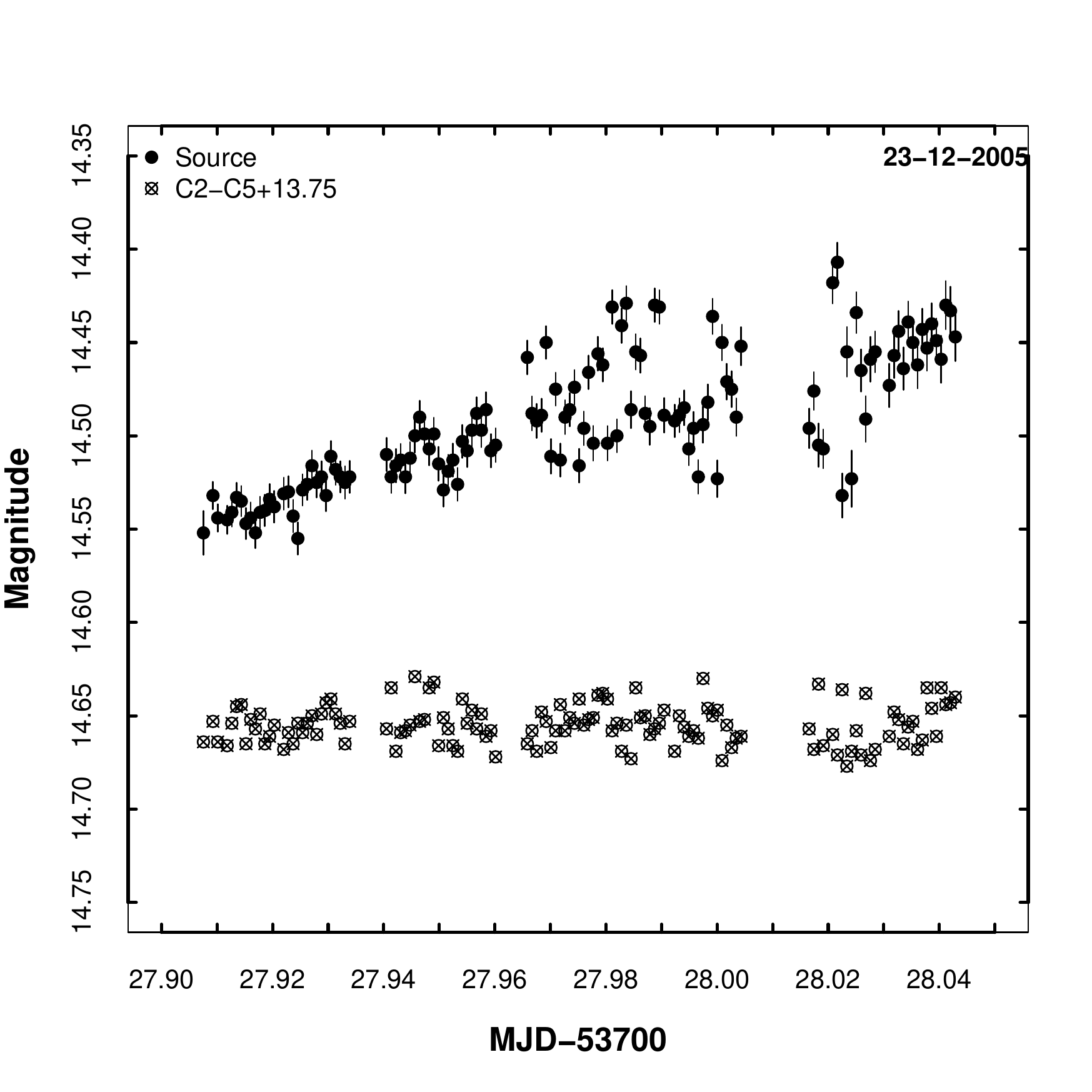}
 \includegraphics[width = 0.3\textwidth]{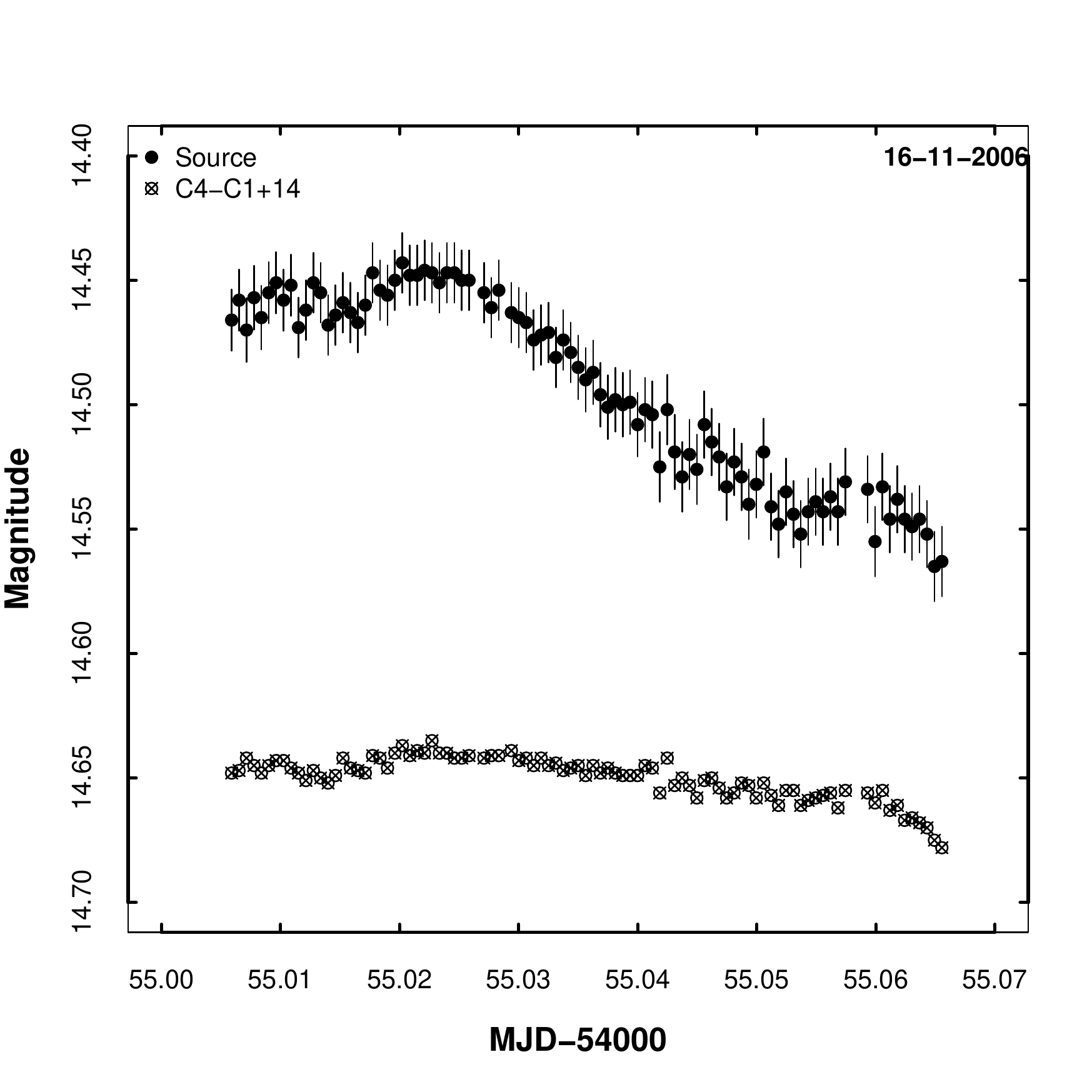}
 \includegraphics[width = 0.3\textwidth]{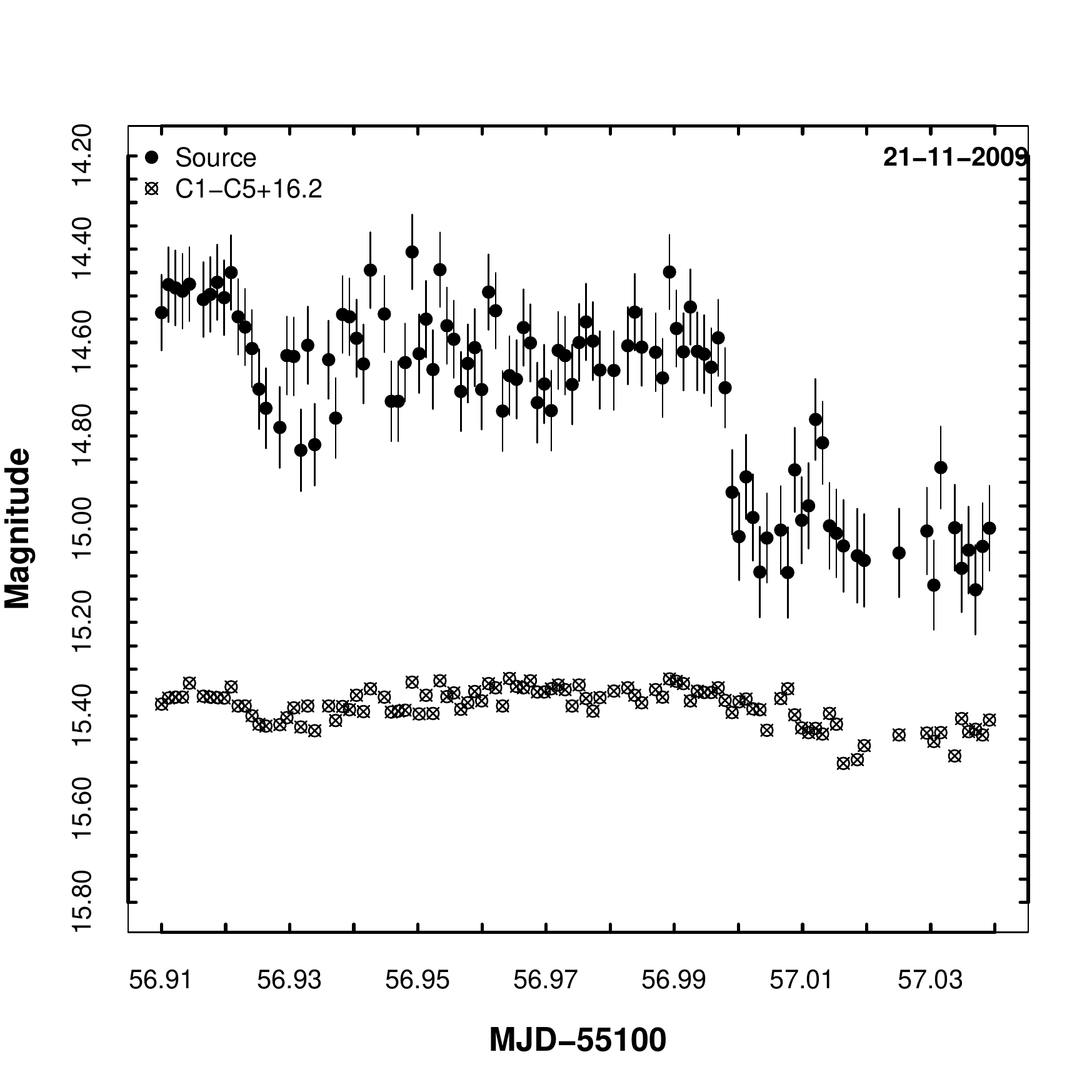}
  \includegraphics[width = 0.3\textwidth]{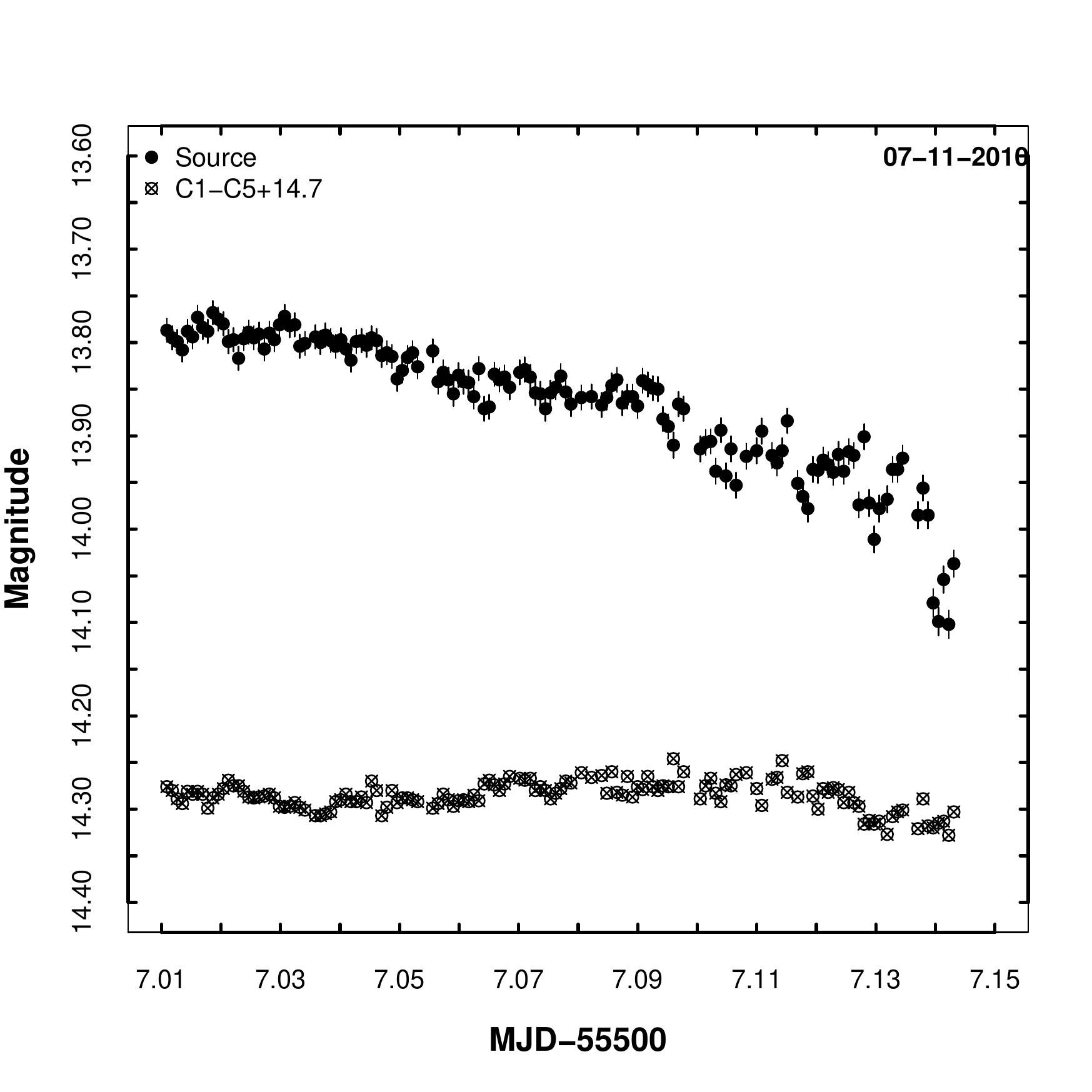}
 \includegraphics[width = 0.3\textwidth]{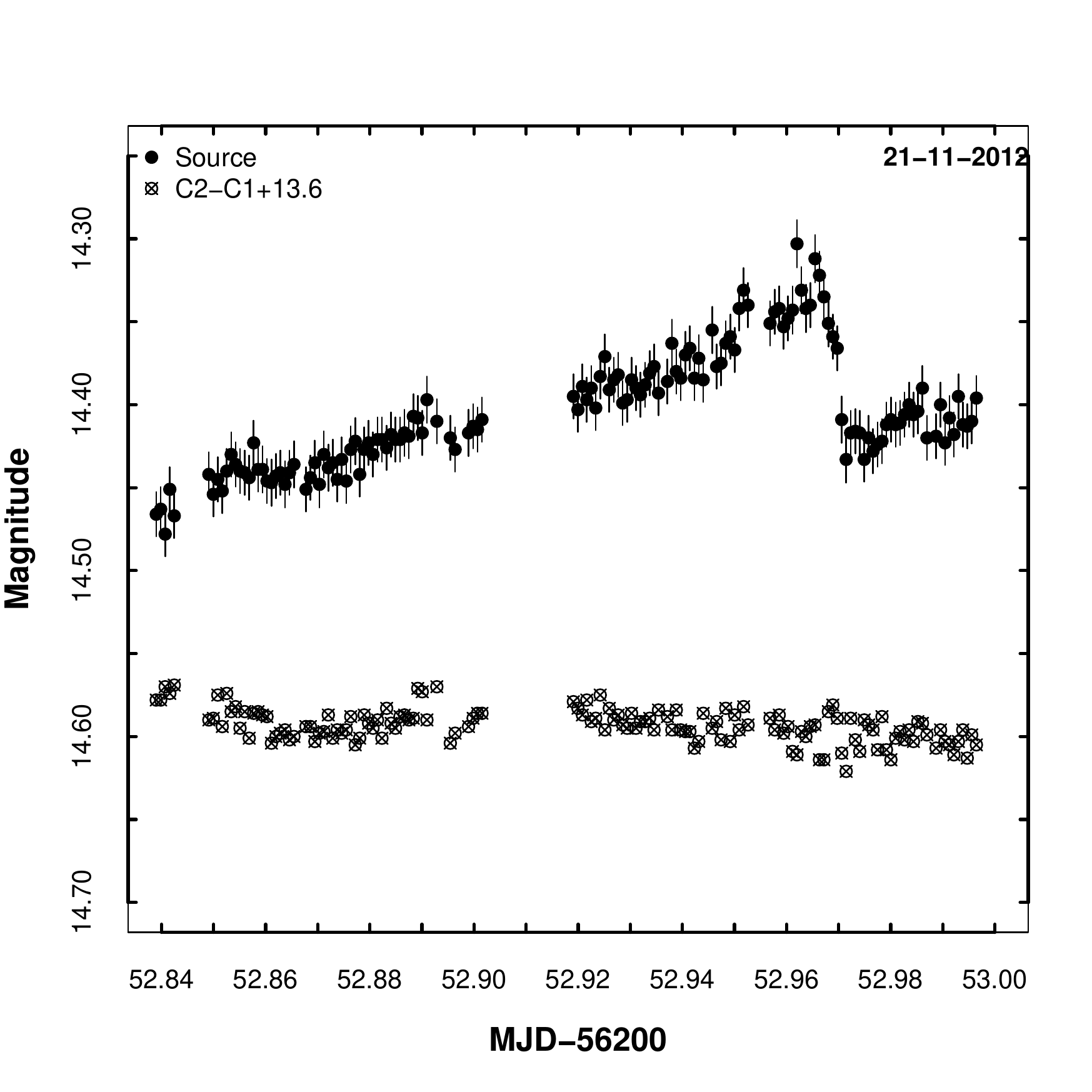}
 \includegraphics[width = 0.3\textwidth]{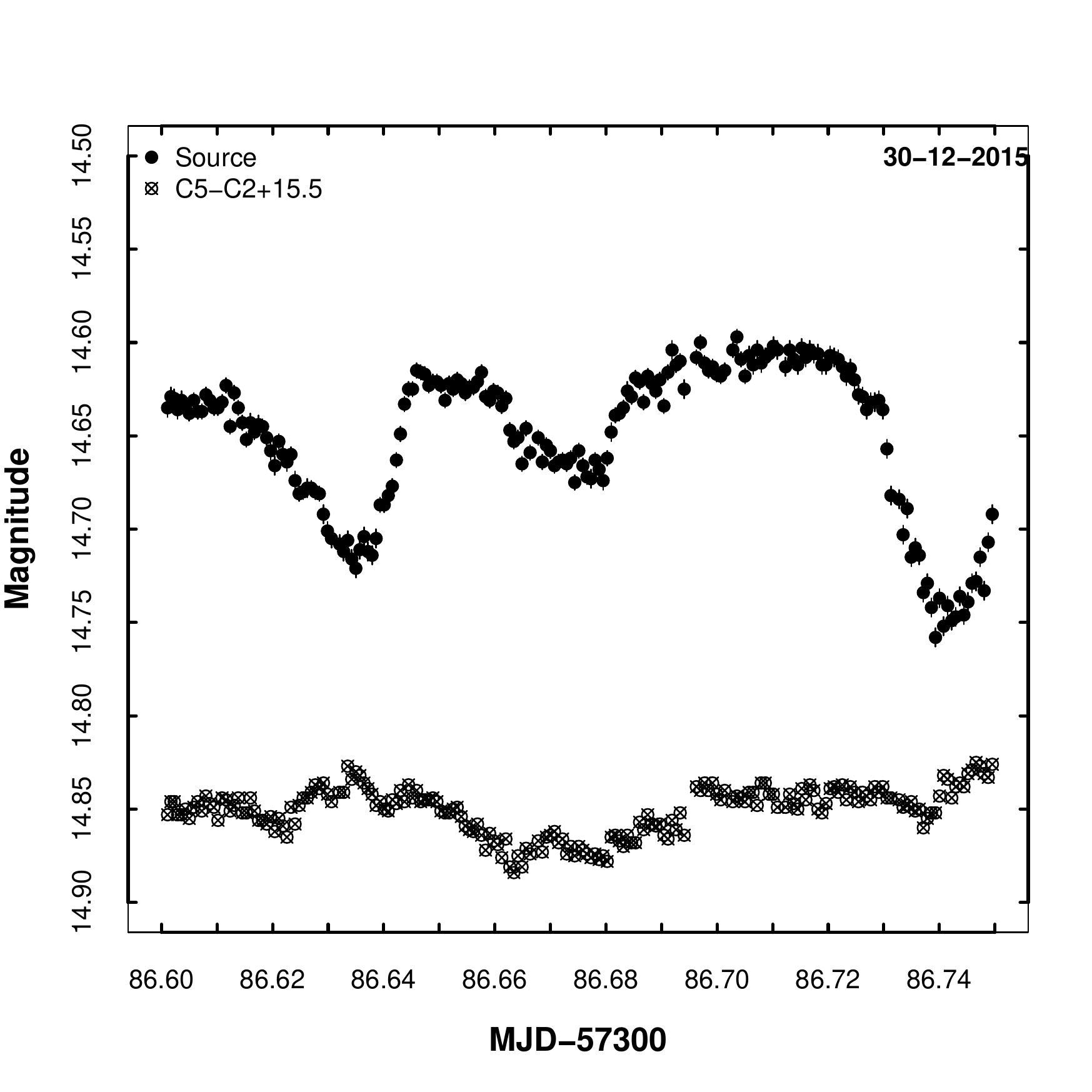}
 \caption{\bf Light-curves for the nights showing optical IDV in 3C\,66A. The statistical results are given in Table \ref{tab:table1}. \label{fig:idvfig}}
\end{figure*}

\subsection{Intra-day Variability (IDV)}
 In order to determine the number of IDV nights, we use the criterion that source should be monitored for at least 1 hr during that night.  We found  89 out of 160 nights to qualify the criterion. Out of these, 13 nights suffered from the bad weather resulting in poor quality light-curves. To verify whether 3C\,66A exhibited IDV during the remaining  76 nights, we apply conditions that the variability parameter C $\geq $2.57, F-value $\geq 5$ and amplitude of variation more than 0.05mag. The nights which satisfied all these criteria were considered as confirmed IDV nights and their observation details along with all estimated parameters  are given in Table \ref{tab:table1}. As can be seen from the Table \ref{tab:table1}, we get only 6 out of the 76 nights which qualify for  confirmed IDV, implying a duty cycle (DC) of variation for 3C\,66A as $\sim$8\% only. However, much higher (up to 86\%) DCs have been reported for 3C\,66A. \citet{Raiteri1998} observed it for two nights and found it variable for both nights (DC=100\%) with 8\% and 14\% amplitude of variation; \citet{Sagar2004}reported IDV on 6 out of 7 nights (DC \textgreater 85 $\% $)  in their 5-10 hrs monitoring during 1998-2002; \citet{rani2011}found the source IDV DC $\sim\ 28\% $ during their 7-night  monitoring for more than 3-hrs while \citet{Gopal2011} observed the source for 5-7 hrs on 7-nights and reported 47\% as duty cycle of variation. We would like to state that the main reasons for our low DC values for 3C66a are (1) relatively shorter duration of monitoring (1 hour and more) compared to 3-10 hrs as reported by these authors, and (2) strict criteria for variability set by us as mentioned above. For example, if we apply the condition of amplitude of variability to be  $\textgreater 5 \% $, DC reported by \citet{Sagar2004} will come down from 86\% to 43\%.  In their studies, \citet{rani2011} \& \citet{Gopal2011} found that longer the period of monitoring, higher the DC for IDV. 

\smallskip

For our 6-IDV nights, amplitude of variation varies between 0.11 to 0.67 mag while C-test parameter lies between 3 to 4.8 and their light-curves are shown in Figure \ref{fig:idvfig}. The IDV light-curves show flux variations which can be broadly classified into following subgroups: (i) a steady rise or fall in the flux during overnight monitoring, (ii) small amplitude fluctuations superimposed over a slowly varying flux and (iii) sharp decline or increase in the flux during the night, giving clear indication of the characteristic time scale of variation. These three types of variations in the light-curves can be attributed to the possibility of the intrinsic differences in the physical mechanisms as well as geometric effects. The continuous rise/fall for a duration of typically few hours may be due to the small scale steady acceleration/cooling of the plasma in the shocked region affecting the synchrotron emission. This can happen if the acceleration/cooling time scales are longer than the light crossing time. Short time scale fluctuations superposed on the slowly varying component might be due to the local inhomogeneities in the blob moving down the jet. The sharp decline or enhancement in the optical flux indicates a violent and evolving nature of the shock formed or sudden injection of the plasma in the emission region. The  timescales shorter than few tens of minutes suggest a possibility that the whole cross-section of the jet may not be shocked. Another reason could be the regions of small scale  over-densities in the blob passing through a standing or slowly moving core \citep{chandra2011}.  Such standing shocks in the jet can be formed by the pressure imbalance between the jet and the ambient medium. Oscillations in the width of the jet occur as the over-pressured matter in the jet expands until its pressure falls below the ambient pressure. This under-pressured plasma then contracts under the influence of external pressure until high pressure is restored. It leads to the formation of an oblique standing shock terminating into a strong shock wave perpendicular to the jet flow \citep{sokolov2004}. In addition,  various plasma instabilities leading to the formation of shocks, magnetic reconnection sites, and turbulence \citep[see the recent discussions in][]{NarayanPiran2012, Subramanian2012, Marscher2014, Saito2015, Sironi2015} can also be responsible for the  IDV and longer time scale variations. Any constraint on these possibilities can only be imposed by using the simultaneous observations at different wavelengths which is beyond the scope of the present study. 

\smallskip

\begin{figure*} 
    \includegraphics[width = 0.30\textwidth]{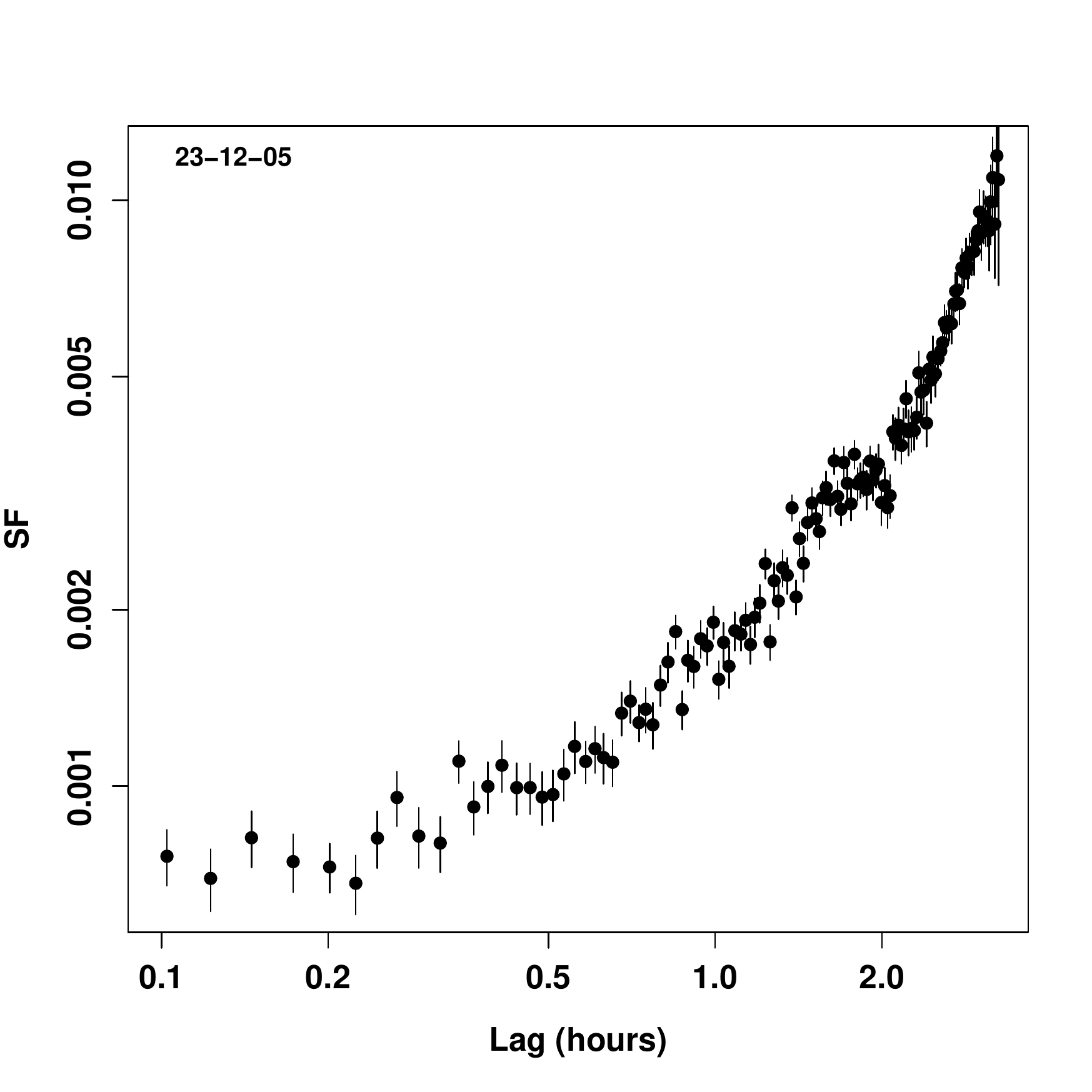}   
       \includegraphics[width = 0.30\textwidth]{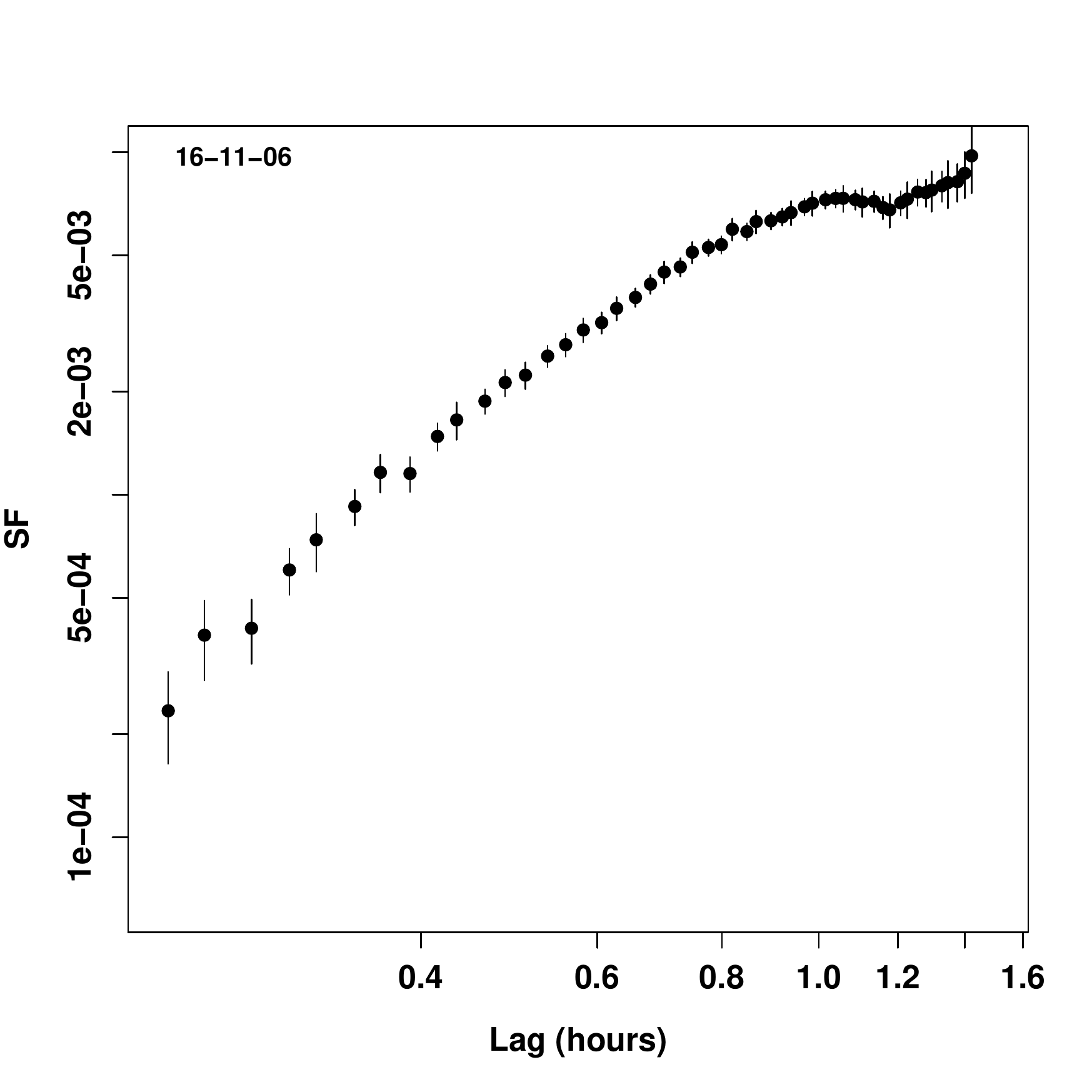}   
    \includegraphics[width = 0.30\textwidth]{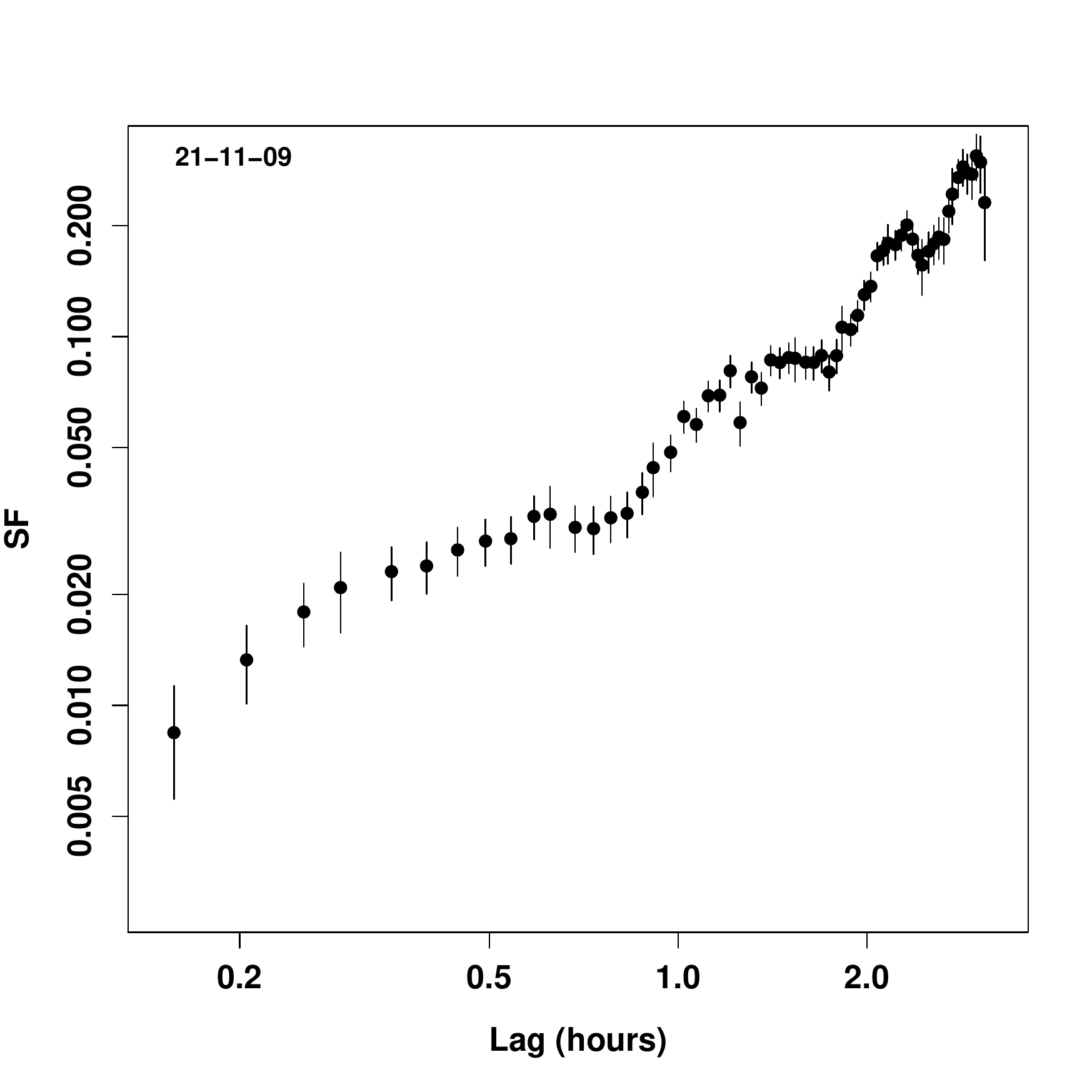}
 \includegraphics[width = 0.30\textwidth]{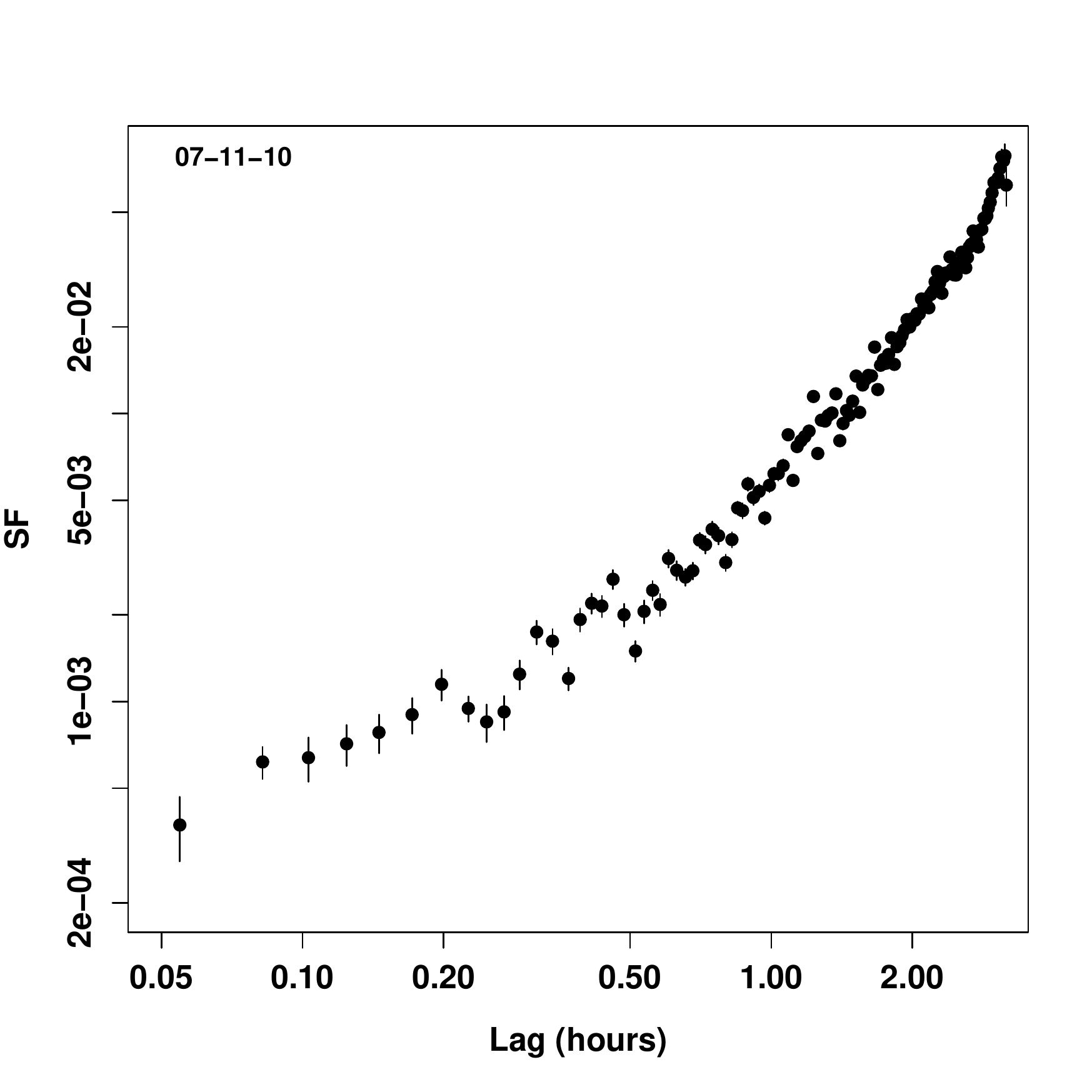}
 \includegraphics[width = 0.30\textwidth]{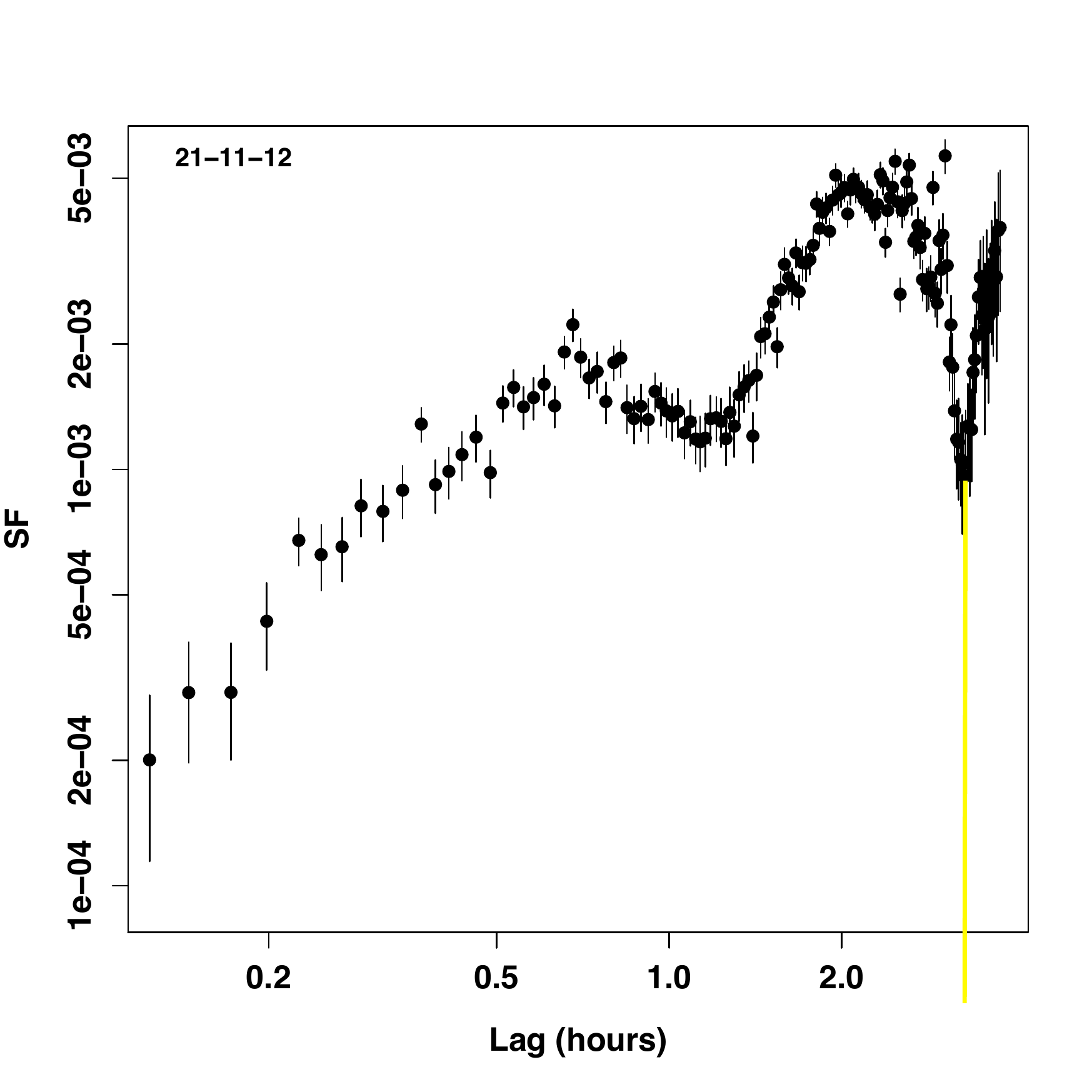}
 \includegraphics[width = 0.30\textwidth]{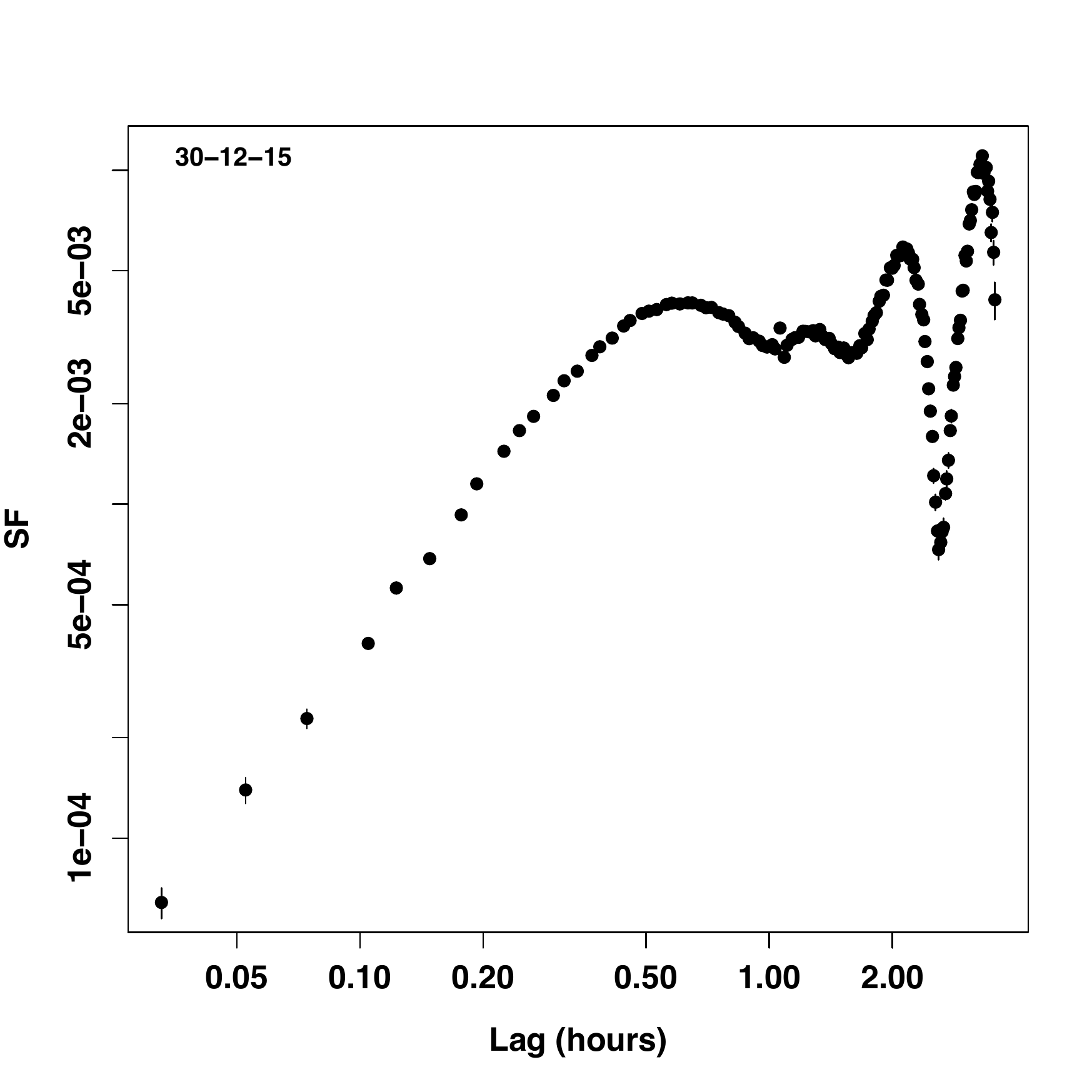}
 \caption{\bf Structure functions for IDV nights of 3C\,66A. Note that 2015 December 30 shows quasi-periodic variations.\label{fig:sffig}}

\end{figure*}

In order to determine various parameters, such as the variability  timescale, extent and the rate of the amplitude of variation etc  in complex cases, structure functions \citep{Simonetti1985, Zhang2012} have been used extensively \citep{carini2011, wu2005, Rani2010, agarwal2016, Dai2015}. We have constructed structure functions (SFs) for all the variable nights in the Figure \ref{fig:sffig}. Using ${\chi}^2$ $-$ minimization method, fits to the SFs were made to calculate timescales and slope of the SFs \citep{Zhang2012} as given in Table \ref{tab:table1}. These are then used to discuss the timescales of variation and the nature of the emission process causing such variations. We notice that the SF slope,  which characterizes the nature of variability, is close to, or more than 1.0 for all the IDV nights, implying that the major cause of the  intra-day variations is the turbulent process \citep{carini2011} at work in the jet as described above.

\subsubsection{Variability time-scales and size of the emission region}

The  shortest time scale of variation puts constraints on the size of the emission region.  The structure functions constructed for the light-curves given in the Figure ~\ref{fig:idvfig},  showed minimal variability timescales in the range of  37 min - 3.1 hrs. Among them, 2009 November 21 (MJD 55156.91) night was found to have the shortest variability timescale of $\sim37$ mins . The light-curve for the night shows indications of multiple small flares almost buried in the noise. The structure function also suggested three plateau and dips but are not  pronounced enough to infer periodicity in the IDV.  For the night of 2015 December 30 (MJD 57386.60), with characteristic variability time scale of 41 minutes, the SF shows signs of  quasi-periodicity and is discussed separately.  
The light-curve for  the  night 2005 December 23 shows increased brightness  with a break in the data at MJD 53728.05 while SF indicated a plateau giving a time scale of 118 minutes. On 2006  November 16, after showing some brightening, the source goes fainter by 0.1 mag. The SF gives a time scale of 68 minutes.  The blazar 3C\,66A goes fainter by about 0.25 mag in R-band in a continuous fall on the night of  2010 November 07. The  structure function shows a monotonic increase without any detectable plateau indicating that the variability timescale  is longer than the duration of the observation.   A continuous increase in the brightness, with a data break at MJD 56252.91 and then sharp drop in brightness by about 0.1 mag characterize the light-curve of  2012 November 21 (MJD 56252.84) while SF gives a variability time scale of about 40 minutes.

\smallskip

Considering the shortest variability timescale, estimated above, as the time required by the light to cross the emitting region, one can set an upper bound to the size of the  emission region, R,  as

\begin{equation}
R \leq \frac {c\delta\Delta t_{min}}{1+z}.
\end{equation}

Taking $\Delta t_{min}$ as 37 mins, $\delta$ as 15 \citep{Bottcher2005} and redshift $z = 0.444$ \citep{Lanzetta1993}, we estimated the size of the emission region to be about  $6.918\times 10^{14}$ cm. Since the variability time scales detected during the 6 IDV nights described here range from 37 minutes to 187 minutes, the estimated sizes of the emission regions vary from $6.92\times 10^{14}$ cm to  $3.496\times 10^{15}$ cm.

The determination of the mass of the central super-massive black holes (SMBH), driving force behind the AGN activity,  is very important for understanding the AGN phenomena. There are several methods to determine the black hole mass but they all require very high spatial resolution to either detect the motion of the stars orbiting the central BH or the detection of the variability in the emission lines in the AGN environment \citep{Chandra2013,Xie2002}. For the BL Lac objects like 3C\,66A, which do not show appreciable emission lines in their spectra, determining BH mass is very difficult. Assuming the origin of the rapid variability in blazar flux from the close vicinity of the black hole, several authors have estimated their mass  using variability time scales \citep[and references there-in] {miller1989, Rani2010, Zhang2012, agarwal2016, Dai2015}. \citet{xie2005} carried out a study to see if the application of the variability timescales was appropriate in estimating the mass of the black hole. They found significant correlation between the masses of the black holes in the blazars  estimated using the variability timescales and those derived from the $L_{5100}-R_{BLR}$ relation. The sample size, however, was  small (12 sources) and hence the inference can not be claimed as conclusive.  Following the  assumption that the observed minimum variability time-scale is manifested by the orbital time-scale of the innermost stable orbit around the central Kerr black hole, we estimate the mass of the black hole using the relation \citep{abramowicz1982, Xie2002}:

\begin{equation}
M=1.62 \times {10}^4 \frac{\delta}{1+z} \Delta t_{min}\mathrm{M}_{\odot}.
\end{equation}

 Using the above relation, we computed the mass of  the black-hole to be 3.7 $\times$ $10^8$ $\mathrm{M}_{\odot}$. This value is in agreement with typical values for such systems. However, as mentioned earlier, such an estimate of the mass of the black hole holds true only when the variations originate very close to the black hole. Therefore, the black hole mass estimated here should be considered with caution.

\subsubsection{Short lived quasi-periodic variations in 3C\,66A }

Several  long term studies of the  blazars have resulted in the detection of the presence of quasi-periodic variations (QPV). In the  case  of OJ287, 12-year quasi-periodicity was explained based on the binary black hole model  \citep{Sillanppa1988, pihajoki2013} while the possibility of the precession of the jet \citep{nesci2005} in S5 0716+714 was suggested as the cause for the alternate decrease and increase of the average source brightness over a period of about 10 years. Though blazars are known for random variations, quasi-periodic variations have been reported at the intra-day time scales in the their optical  light-curves \citep{gupta2009, wu2005, Rani2010}. \citet{Rani2010} detected QPV in optical with a time scale of 15 mins. \citet{wu2005} explained presence of a sine-like feature in the intra-day light-curve of S5 0716+71 based on light-house effect.  

In the present case, out of the 6-IDV nights, light-curve and structure function for the night of 2015 December 30 show a possibility of the QPV. In the structure function analysis, for the time scale of variation, we have taken the interval of time from where the SF begins a steady ascent to its attaining a peak (slope vanishes to zero). For the periodic variation, if there are more than one minima in the SF, the interval between consecutive minima is taken as the time period. When the intervals between several such minima are almost similar, their average value is taken as the time scale for the QPV. If the interval values turn out to be much different, periodic variation is questionable or there might be more than one time scale present in the system. The SF for the night of 2015 December 30 (see Figure ~\ref{fig:sffig}) shows maximum at 0.65 hr, first minimum at 1.07 hrs, second at 1.56 hrs and the third one at 2.59 hrs giving the intervals or timescales as 1.087 hrs, 0.48 hr and 1.03 hrs, respectively. There appears to be two distinct time scales of  1.05 hrs and 0.48 hr in this case, as one value is very different from the rest of the two values. It, therefore, indicates to the possibility of  quasi-periodic variations during this night, also reflected in the light-curve for that night. However, such time scales appear only in a part of the light curve and could be transient in nature.

In order to cross check the SF results we used  L-S periodogram\citep{Lomb1976, Scargle1982}  and discrete correlation function (DCF:\citet{Edelson1988, Hufnagel1992, Alexander2014}) methods to ascertain the existence of QPVs. For this,  in order to remove slowly varying component from the light-curve of 2015 December 30, we subtracted a linear fit. The statistical technique of DCF and L-S periodogram was applied to the resulting data.

\smallskip

\begin{figure} 
 \includegraphics[width = 0.50\textwidth]{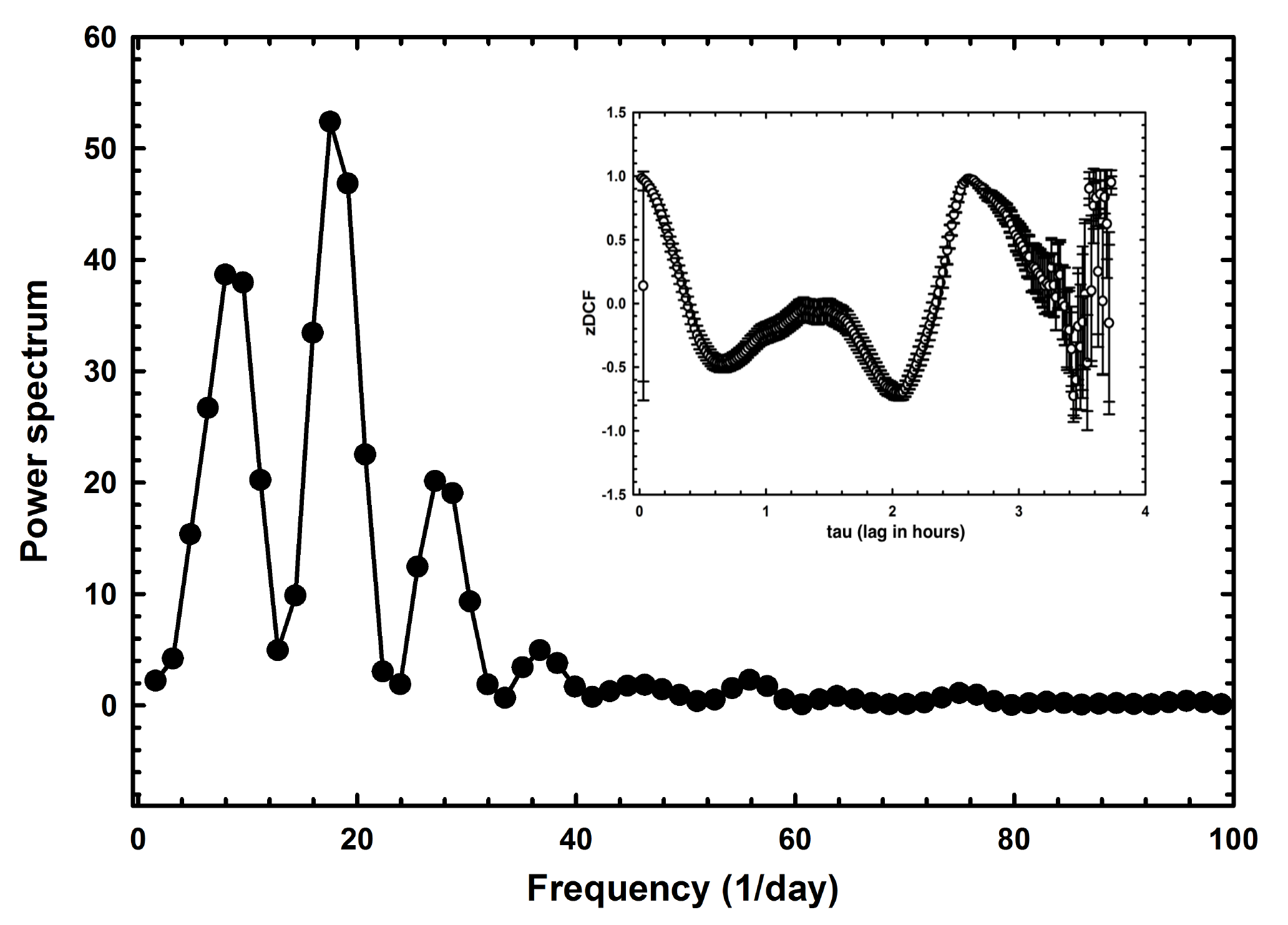}

 \caption{\bf  LSP and zDCF (inset) for the 2015 December 30 IDV night. The peaks in LSP and zDCF  at more than 99\% significance level indicate to the possibility of QPV with  a quasi-periodic component of about 1.4hr.\label{fig:qpo_case}}

\end{figure}

\smallskip

On 2015 December 30 (MJD 53726.99), the periodogram reveals most significant peak at  1.4 hrs with DCF also giving a peak separated by a time lag of about 1.4 hrs. There is another peak in the power spectrum and zDCF showing a period of about 2.65 hrs but that is close to the length of the data series (duration of observations). These values support the suggestion from structure function indicating to the presence of quasi-periodic variations.  These can be caused by the micro-lensing, a geometrical effect, \citep{Watson1999, Webb2000} which leads to symmetric and  periodic features in the light-curves which are achromatic in nature. Also, time scales should be of the order of a day or longer. As indicated by our light-curves, these short time-scale features are not symmetric in nature and hence are not expected to be due to micro-lensing effect. Shock-in-jet model \citep{MarscherGear1985} can not directly explain these features.  \cite{Camenzind1992} invoked a geometric light-house kind of approach to explain such quasi-periodic behaviour in the frame work of shock-in-jet model.  According to it,  these intrinsic quasi-periodic variations  are caused by the plasma streamlet moving down the jet in a helical motions.

\subsubsection{Rise and fall rates of flare brightness}

The IDV light-curves shown in Figure ~\ref{fig:idvfig}, reflect smooth rise or fall  in the source brightness over the night.  Three  nights show slow rise/fall in flux superposed by rapid fluctuations. The rate of change of the flux at different epochs is determined by fitting a line segment to the light-curves. The mean value of  the rising rates  is approximately 0.06$\underline{+}$0.02 mag hr$^{-1}$. The fastest rate of the change in optical flux  (1.38mag/hr) has been seen on 2009 November 21 (MJD 55156.99) when the source decays by 0.40 mag within about 17 mins. Though the error bars are large, trend is very clear (more than $3 \sigma$ change). Almost a smooth brightening, at a rate of 0.03mag/hr and smooth decline, at the rate of 0.078mag/hr are noticed in the light-curves of 2005 December 23 and 2010 November 7, respectively. Several changes in the source magnitude are seen on the night of 2015 December 30, the night showing QPV, along with a flat trend for a period of about 55 minutes. However, some of the flares on this night lead to rapid changes, varying from 0.12 mag/hr to 0.41mag/hr (errors are within the data points drawn). None of these features appear to be symmetric in nature or having similar rates of flux  changes while rising and falling. The flux evolution in such cases is probably not caused by the light crossing time. If the particle injection and cooling were operating at a time scale shorter than R/c (R, size of the emission region), we expect symmetrical rise and fall profiles. Therefore, either injection/acceleration or cooling time scale must be affecting variation, causing different rise and fall rates.

\subsubsection{Brightness state of the source and the extent of variation}

In order to investigate the variability activity in different phases of the source brightness, we plotted  variability amplitudes, estimated for all the nights (76 nights) considered for the intra-day variability test,  as a function of their respective  daily mean brightness magnitudes.  
The nature of the  plot  revealed  larger intra-day variability amplitudes during the nights when the source is relatively fainter. Normally, one would expect  larger amplitude of variation during the bright, flaring phases of the source when highly turbulent jet plasma is expected to interact with the frequent shock formations leading to its rapid acceleration and subsequent radiative cooling. During the quiescent phase, the jet is relatively quiet and the blazar emission might have some contribution from other components, for example accretion disk, BLR and host galaxy thermal emission. However, it has been seen that while flat spectrum radio quasars (FSRQ) show contribution from these sources of emission, BL Lac objects show much less such contribution to the optical flux which is dominated by  the beamed  emission from the jet \citep{Raiteri2012}, even when the source is in the low phase of brightness.  One would, therefore, not expect variations with large amplitude  during the nights when the source is in low state of brightness. Perhaps, during the bright phase, amplitude of variation is the victim of the overlapping effects of a large number of rapid flares, resulting in the  higher base level brightness and reduced effective amplitude of the variation. On the other hand, if we drop  a few outliers in the plot, trend would seem to be almost  independent of the state of brightness. That is what one expects if the variability is caused by the processes in the jet.  In any case, in order to make a definitive statement on this trend, more such data on a sample of blazars are required.

\subsection{Long-term Variability of the blazar 3C\,66A}

\subsubsection{Average brightness trend during 2005$-$2015}

\begin{figure} 
 \includegraphics[width = 0.5\textwidth]{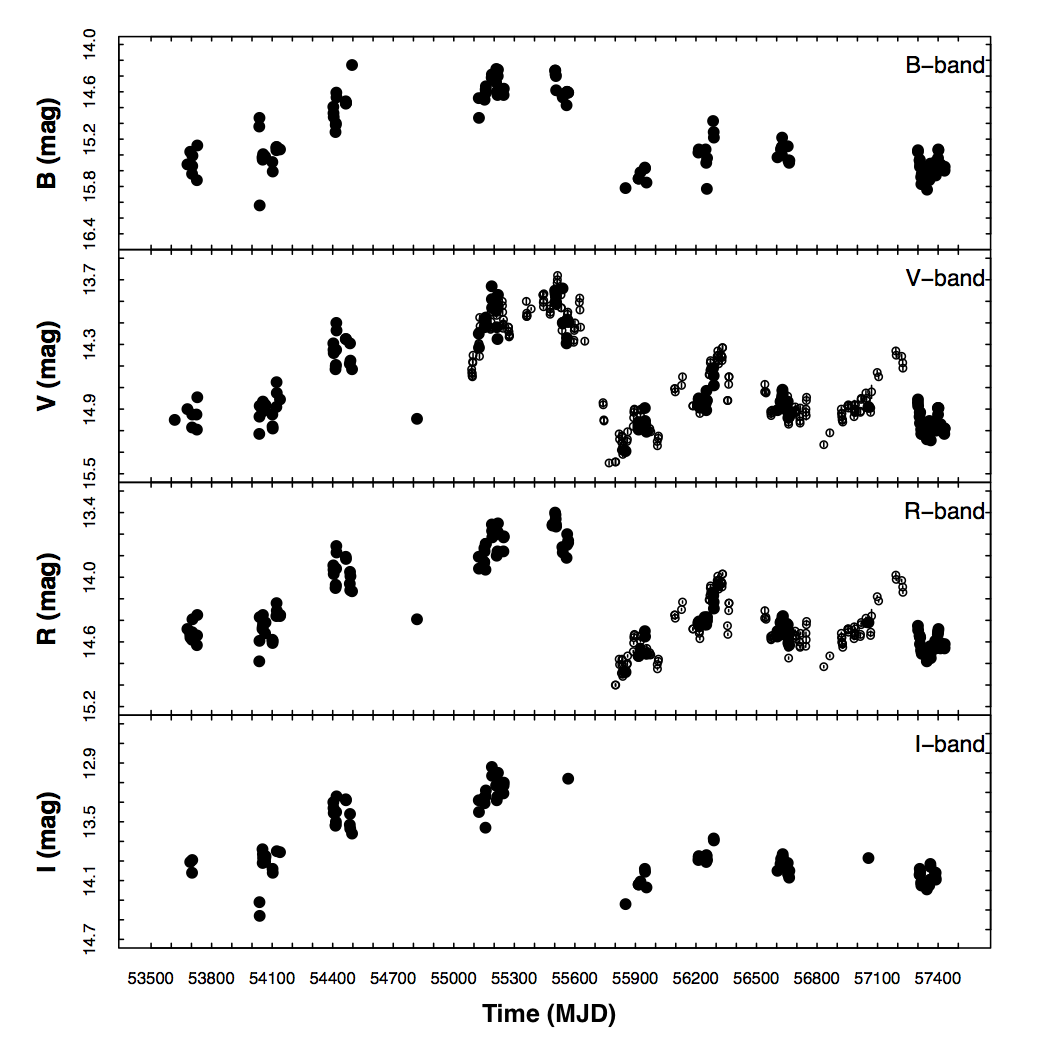}
 \caption{\bf Multi-band optical light-curves  for 3C\,66A during 2005-2016.  Steward Observatory data in R \& V bands  are shown with open circles.\label{fig:lt_bvri}}
 \end{figure}

To understand the long term behavior of the source and to determine any long-term variability pattern, we used the data for all the nights during 2005 to 2016, averaging the magnitudes on the daily basis and plotted the light-curves for all the optical bands (BVRI). We have also used the archival data of the Steward observatory \citep{Smith2009}  available online in R and V bands, to fill the gaps in our daily averaged  data set. The R-band light-curve (see 3rd panel from top in Figure~\ref{fig:lt_bvri}) thus plotted shows several flares, superposed over the three outbursts- one major (during MJD 53680 to MJD 56000) peaking somewhere around MJD 55501 (2010 November 1) with R=13.40 mag and two minor ones, decaying and rising parts of which could  be overlapping. These two outbursts  peak  at about MJD 56272 and MJD 57192 with almost similar brightness values,  14.08 and 14.02 mag, respectively.  From the light-curve one can notice the  brightness changing by about 0.6 mag within a period of tens of  days (STV), by about 1.3 mag in about 700 days and by 1.7 mag within a span of about 1650 days (LTV). There are evidences of the shorter term variations with smaller brightness changes.  Thus the light-curves show presence of long-term and short-term variabilities (LTV \& STV) during the period of our observation campaign. However, due to the lack of the continuous data set, we are not able to fully resolve and trace these flares and outbursts to be too quantitative. In the course of our decadal observation campaign, 3C\,66A has shown a waxing trend reaching its peak brightness (13.40 mag in R band) on 2010 November 1 (MJD 55501.04) and then a waning trend towards later half of the decade. A definite decline in the amplitude of variations in the later half of the decadal observation campaign is noticeable. Similar pattern is reflected in the other optical bands  B, V and I as also shown in the figure. 

During  the course of the campaign the average R band magnitude was 14.10 mag, with a maximum as 13.4 mag on 2010 November 1 (MJD 55501.04) and a minimum of 15.01 mag (indicated by the Steward data point in Figure 5) on 2011 August 27 (MJD 55800). The brightness magnitudes for the other bands are as follows (maximum$-$average$-$minimum): B-band (14.10$-$15.10$-$16.24), V-band (13.41$-$14.45$-$15.85) and I-band (12.90$-$13.49$-$14.77).

A natural explanation for the flux fluctuations in the blazars, at both long and short term time scales, is the presence of turbulence in the relativistic jet that emits  non-thermal synchrotron radiation \citep{Marscher2014,Romero1999}. STVs with timescales of the order of days could be a result of the helical jet model \citep{MarscherTravis1996}. The source brightness varies with the viewing angle: the source is brighter when the viewing angle is small, but fainter when the viewing angle is large \citep{Lainella1999}. However, perturbations and or instabilities in the accretion disk could also lead to such variations on longer time scales indicated in our light-curves. 

We constructed L-S periodogram from the R-band light-curve data to check whether any quasi-periodic variation exists in the long-term data series. We find significant ($>$99\% confidence) power on a time scale of 2.48 years. DCF of the same data set also shows peaks separated, roughly,  by a time interval of about 2.50 years which may be taken as a possible characteristic time scale. Perhaps every 2.5 years or so, a major flare occurs in the jet, caused by a fresh injection of the plasma in the relativistic jet of the source 3C\,66A. More data are required to arrive at a better analysis of the long term behavior of the blazar 3C\,66A as there are several large gaps in the data series, even after including the R-band data from the Steward Observatory.  

\smallskip
Considering the issue of  whether  IDV and LTV are related in some way as a result of  the processes in the relativistic jet, blazars are the best candidate to study such relationship. Since IDV is mainly caused by the interaction of small scale inhomogeneities with the plasma in motion resulting in the enhancement of the synchrotron emission, and/or injection of fresh plasma in the jet, it can very well contribute to the long-term variability. Certainly these processes enhance the jet emission but it is more of a local effect than global. Perhaps it would be meaningful to find LTV time scales of variation and try to see if there was any correlation between short term and long term timescales of the variability. However, for this a well sampled data set is required to un-ambiguously determine the two timescales. In our study, IDV amplitudes are inversely correlated to the brightness of 3C\,66A but the correlation is not very strong as there are several outliers in the plot. The role of the disk in the long term variability (LTV) is understandable but when it comes to the rapid variations, processes in the jet are the most credible source of the variation. Certainly more study is required to address the interesting issue of a relationship between the short and the long term variations.

\subsubsection{Long-term spectral behavior of 3C\,66A}

\begin{figure}
\includegraphics[width = 0.50\textwidth]{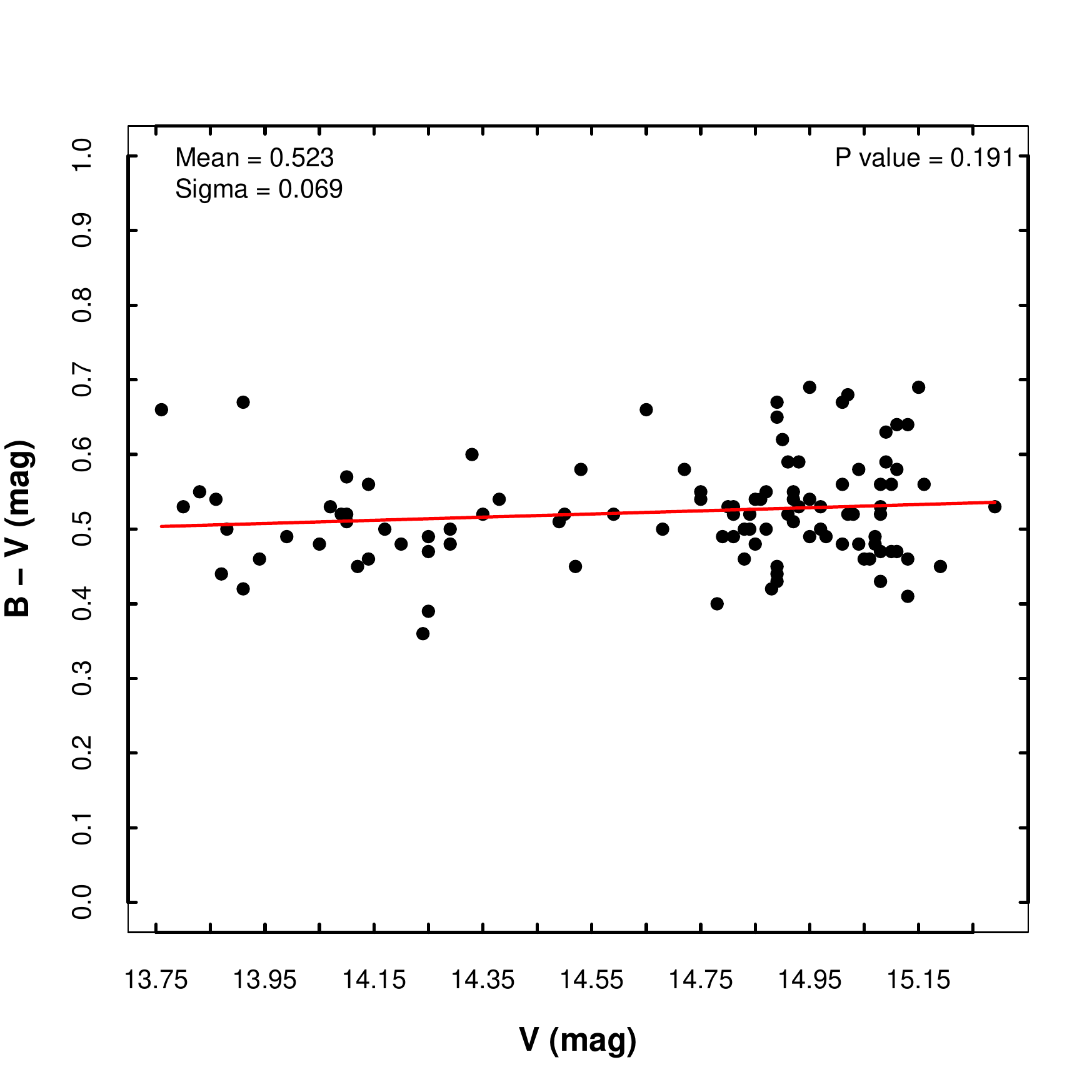}
\caption{\bf Color as a function of the brightness of the source 3C\,66A, straight line shows best fit to the data.\label{fig:color_state}}
\end{figure}

The data obtained in BVRI bands for all the nights of our 10$+$ year campaign, irrespective of the duration of observation on a particular night, is used to understand the spectral behavior of the source. The color information is important as it helps to discriminate among the various physical processes responsible for the variation. Using the observed data in B, V, R, and I bands for all the nights from 2005 November 06 to 2016 February,  each night's averaged colors (B-I, B-V, R-I, V-R) are determined. We have plotted (B-V) color as a function of the V band magnitude (see Fig. \ref{fig:color_state}) using whole dataset  which shows  a mild bluer-when-brighter (BWB) trend which is typical of BL Lac objects. The similar trend is seen in the plots for (B - I)  v/s I, (B - R) v/s R etc with a marginal difference in the slope. The bluer when brighter trend  could be explained based on the shock$-$in$-$jet model \citep{MarscherGear1985}.  Also, it can occur when an increase in the luminosity of the blazar is caused by the injection of a population of fresh electrons/plasma with a harder energy distribution than the older partially cooled ones  \citep{Mastichiadis2002}.  We have also studied the behaviour of the color as a function of the time during the 10 year period presented here. We notice a very weak trend of decrease in the color with time.

During the  WEBT-ENIGMA multi-wavelength campaign in 2004, the BL-Lac object 3C\,66A was extensively monitored which  gave a hint of the spectral evolution during an optical flare  \citep[see][]{Bottcher2005}. The source showed a very weak positive correlation between the (B-R) color index and the R-band magnitude while in a low phase. A similar trend of the  positive correlation was reported by \citet{Vig2003} and \citet{Gu2006}. Several studies indicated a significant spectral variation exhibited by 3C\,66A when in the low-flux phase. However, there was no hint of any spectral variation during the high-flux phase (R$<$13.5) of the source \citep{Bottcher2009, Rani2010}.

To understand the spectral variation as a function of the average brightness of the source, we categorized source to be in high state when R $\leq 14.0$ and in low state when R $>$ 14.0. Using the Pearson's product moment correlation, we computed correlation and the significance of correlation in the low and high states for V-R and B-V. Our results show almost similar correlations in both the cases, whether source is in the low or high state of brightness. Both show a mild bluer when brighter trend.  
Generally, blazars are found to follow a common trend of getting bluer when in brighter phase, but one cannot rule out the observational evidence for a few blazars getting redder when brighter  \citep{stalin2006}. This scenario really needs a careful look on the color index, spectral slope and brightness of the source before drawing any general conclusion about the category of the blazars. Also, \citet{Raiteri2003} reported the different spectral behaviors found at shorter timescales while \citet{Gu2006} found no correlation between color vs brightness of the source during the IDV nights. There are studies on the  color behavior of a sample of FSRQs and BL-Lacs \citep[e.g.][]{Raiteri2012}. Their study distinctly shows a RWB color for the FSRQs and BWB trend for the BL Lacs, in general. 

\section{Conclusions}
 Long term high spatial resolution data are presented for the duration from 2005 November to 2016 February for the blazar 3C\,66A.  Nightly averaged B, V, R and I magnitudes for the whole duration obtained from MIRO along with  R and V band data from Steward Observatory, where available, are used to address the long term variability and color behaviour of the blazar 3C\,66A during this period.   Based on these observational results, following conclusions are drawn:
 \\
%\begin{itemize}
1. The duty cycle for the intra-day variability in 3C\,66A is about 8\%, which is relatively lower for BL Lacs, perhaps due to short duration monitoring.
\\
2. The variability time scales for the IDV vary from about  37  minutes to about 3 hrs indicating to different sizes for the emission regions. The shortest time scale of variation leads to an upper limit on the size of the emission region as $6.92\times10^{14}$ cm. Consequently, the mass of the central SMBH is estimated to be about $3.7\times10^8 \mathrm{M}_{\odot}$, assuming the origin of the variability from the region very close to the central engine.
\\
3. Short-lived quasi-periodic variations (QPVs) are probably present in one IDV night with a time scale of about 1.4 hrs.  This could be the result of the plasma streamlet progressing down the jet in a helical motion, described under the shock-in-jet model.
\\
4. Long term trend in the source brightness shows a large number of flares superposed on the slowly varying pattern. There is an indication of a possible characteristic timescale of about 2.5 years, however large gaps in the data forbid us to state it confidently.
\\
5. A slowly decreasing trend in the average brightness with time and a mild bluer when brighter behavior are noticed, the later supports the shock-in-jet model, typical for BL Lacs. 
%\end{itemize}

 The results of the study are drawn from a data set which has large gaps and hence the statistical analysis could be having a bias. However, we believe that it is the largest data set on 3C\,66A and would be very useful for the community. Nonetheless, there is a great need for having better data sampling, perhaps using all the published data and more co-ordinated observations to appropriately address the issues of  intra- and inter-night variations and the physical processes responsible for them. 

\section{Acknowledgements}

This work is supported by the Department of Space , Government of India. We acknowledge the help of Mr. Kumar Venkatramani, Dr. Sunil Chandra,  the staff of the Mt. Abu InfraRed Observatory for help in observations.  We appreciate the anonymous referees for their constructive remarks leading to a much improved manuscript. We also acknowledge the use of the data from the Steward Observatory spectropolarimetric monitoring project, supported by the Fermi Guest Investigator grants NNX08AW56G, NNX09AU10G, NNX12AO93G, and NNX15AU81G.
\bibliography{reference}

\end{document}